\documentstyle[12pt]{article}
\def\newpic#1{%
   \def\emline##1##2##3##4##5##6{%
      \put(##1,##2){\special{em:point #1##3}}%
      \put(##4,##5){\special{em:point #1##6}}%
      \special{em:line #1##3,#1##6}}}
\newpic{}

\newtheorem{theorem}{Theorem}[section]
\newtheorem{lemma}[theorem]{Lemma}
\newtheorem{corollary}[theorem]{Corollary}
\newtheorem{proposition}[theorem]{Proposition}
\newtheorem{remark}[theorem]{Remark}
\newtheorem{example}[theorem]{Example}
\newcommand{\bull}{\mbox{$\;\;\;$\vrule height .9ex width .8ex depth -.1ex}}
\newenvironment{proof}{\par\smallbreak\noindent{\bf Proof.~}}%
{\unskip\nobreak\hfill \bull \par\medbreak}
{\unskip\nobreak\hfill \bull \par\medbreak}
\newcommand{\case}[2]{{\it Case #1:\/ #2.}}
\newcommand{\subcase}[2]{\par{\it Subcase #1:\/ #2.}}
\newenvironment{venumerate}%
{\begin{enumerate}}%
{\end{enumerate}}
\newcounter{oq}
\newcommand{\que}{\refstepcounter{oq}\par{\bf \theoq.}~}

\newcommand{\refeq}[1]{(\ref{#1})}
\newcommand{\function}[2]{:#1 \rightarrow #2}
\newcommand{\of}[1]{\left( #1 \right)}
\newcommand{\setdef}[2]{\left\{\hspace{0.5mm}#1:\hspace{0.5mm} #2\right\}}

\newcommand{\gs}{\sigma}
\newcommand{\om}{\omega}
\newcommand{\gh}{\gamma}

\newcommand{\force}[1]{F_\chi( #1 )}

\newcommand{\numcol}[1]{N_\chi( #1 )}
\newcommand{\fforce}[1]{F^*_\chi( #1 )}
\newcommand{\decforce}[1]{\mbox{{\sc Force\/}${}_\chi( #1 )$}}
\newcommand{\decfor}[2]{\mbox{{\sc Force\/}${}_{#1}( #2 )$}}
\newcommand{\cccol}{\mbox{\rm 3COL}}
\newcommand{\ucol}{\mbox{\rm U3COL}}
\newcommand{\sat}{\mbox{\rm SAT}}
\newcommand{\usat}{\mbox{\rm USAT}}
\newcommand{\cloUS}%
{\mbox{P}^{\mbox{\scriptsize US}}_{\mbox{\scriptsize dtt}}}
\newcommand{\pnplog}{\mbox{P}^{\mbox{\scriptsize NP[$\log n$]}}}
\newcommand{\mone}%
{\mathbin{{\le}_{\mbox{\scriptsize m}}^{\mbox{\scriptsize P}}}}
\newcommand{\disj}%
{\mathbin{{\le}_{\mbox{\scriptsize dtt}}^{\mbox{\scriptsize P}}}}
\newcommand{\turi}%
{\mathbin{{\le}_{\mbox{\scriptsize T}}^{\mbox{\scriptsize P}}}}
\newcommand{\opt}{\mathop{\rm opt}\nolimits}
\newcommand{\sol}{\mathop{\sl sol}\nolimits}
\newcommand{\optsol}{\mathop{\sl optsol}\nolimits}
\newcommand{\I}{\mbox{\sl I}}
\newcommand{\val}{\mbox{\sl v}}
\newcommand{\UO}{\mbox{\rm UO}}
\newcommand{\US}{\mbox{\rm US}}
\newcommand{\cpclass}{\mbox{\rm C${}_{=}$P}}
\newcommand{\ppclass}{\mbox{\rm PP}}
\newcommand{\forcef}{\mathop{\it force}\nolimits}
\newcommand{\sitwo}{\Sigma_2^{\mbox{\scriptsize P}}}
\newcommand{\calY}{{\cal Y}}

\newif\ifnotesw\noteswtrue
\newcommand{\comm}[1]{\ifnotesw $\ll${\sf #1}$\gg$\fi}
\noteswfalse	

\title{On the Computational Complexity of\\
the Forcing Chromatic Number}

\author{Frank~Harary\thanks{Computer Science Department,
New Mexico State University,
Las Cruces, NM 88003, USA.}\quad
Wolfgang~Slany\thanks{Institut f\"ur Softwaretechnologie,
Technische Universit\"at Graz,
A-8010 Graz, Austria.}\quad
Oleg Verbitsky\thanks{%
Institut f\"ur Informatik,
Humboldt Universit\"at Berlin,
D-10099 Berlin, Germany.
Supported by an Alexander von Humboldt fellowship.}}

\date{30 August 2005}

\begin{document}
\maketitle

\begin{abstract}
We consider vertex colorings of graphs in which adjacent vertices
have distinct colors. A graph is $s$-chromatic if it is colorable in $s$
colors and any coloring of it uses at least $s$ colors.
The forcing chromatic number $\force G$ of an $s$-chromatic graph $G$ is
the smallest number of vertices which must be
colored so that, with the restriction that $s$ colors are used, every
remaining vertex has its color determined uniquely.
We estimate the computational complexity of $\force G$
relating it to the complexity class US introduced by Blass and Gurevich.
We prove that recognizing if $\force G\le2$ is US-hard with respect
to polynomial-time many-one reductions. Moreover, this problem is coNP-hard
even under the promises that $\force G\le3$ and $G$ is 3-chromatic.
On the other hand, recognizing if $\force G\le k$, for each constant $k$,
is reducible to a problem in US via a disjunctive truth-table reduction.

Similar results are obtained also for forcing variants of the clique
and the domination numbers of a graph.
\end{abstract}

\section{Introduction}\label{s:intro}

The vertex set of a graph $G$ will be denoted by $V(G)$.
An {\em $s$-coloring\/} of $G$ is a map from $V(G)$ to $\{1,2,\ldots,s\}$.
A coloring $c$ is {\em proper\/} if $c(u)\ne c(v)$ for any adjacent vertices
$u$ and $v$. A graph $G$ is {\em $s$-colorable\/} if it has a proper
$s$-coloring. The minimum $s$ for which $G$ is $s$-colorable is called
the {\em chromatic number\/} of $G$ and denoted by $\chi(G)$. If
$\chi(G)=s$, then $G$ is called {\em $s$-chromatic}.

A {\em partial coloring\/} of $G$ is any map from a subset of $V(G)$
to the set of positive integers. Suppose that $G$ is $s$-chromatic.
Let $c$ be a proper $s$-coloring and $p$ be a partial coloring of $G$.
We say that $p$ {\em forces\/} $c$ if $c$ is a unique extension of $p$
to a proper $s$-coloring. The domain of $p$ will be called a {\em defining
set for $c$}. We call $D\subseteq V(G)$ a {\em forcing set in $G$\/}
if this set is defining for some proper $s$-coloring of $G$. The minimum
cardinality of a forcing set is called the {\em forcing chromatic number\/}
of $G$ and denoted by~$\force G$.
This graph invariant was introduced by Harary in~\cite{Har}.
Here we study its computational complexity.

To establish the hardness of computing $\force G$, we focus
on the respective slice decision problems which are defined for
each non-negative integer $k$ as follows:

\smallskip

\decforce{k}

{\it Given:} a graph $G$.

{\it Decide if:} $\force G\le k$.

\smallskip

\noindent
The cases of $k=0$ and $k=1$ are tractable. It is clear that $\force G=0$ iff
$\chi(G)=1$, that is, $G$ is empty. Furthermore, $\force G=1$ iff
$\chi(G)=2$ and $G$ is connected, that is, $G$ is a connected bipartite
graph. Thus, we can pay attention only to $k\ge2$. Since there is
a simple reduction of \decforce{k} to \decforce{k+1}
(see Lemma \ref{lem:red1}), it would suffice to show that even \decforce{2}
is computationally hard. This is indeed the case.

Let \cccol\/ denote the set of 3-colorable graphs and \ucol\/ the set
of those graphs in \cccol\/ having a unique, up to renaming colors,
proper 3-coloring. First of all, note that a hardness result for
\decforce{2} is easily derivable from two simple observations:
\begin{equation}\label{eq:simple}
\parbox{70mm}{%
If $\force G\le2$, then $G\in\cccol$;\\
If $G\in\ucol$, then $\force G\le2$.}
\end{equation}
The set \cccol\/ was shown to be NP-complete at the early stage of
the NP-completeness theory in \cite{Sto,GJS} by reduction from
\sat, the set of satisfiable Boolean formulas.
It will be benefittable to use a well-known stronger fact:
There is a polynomial-time many-one reduction $p$ from \sat\/
to \cccol\/ which is {\em parsimonious}, that is, any Boolean formula
$\Phi$ has exactly as many satisfying assignments to variables as the
graph $p(\Phi)$ has proper 3-colorings (colorings obtainable from one another
by renaming colors are not distinguished). In particular, if $\Phi$ has
a unique satisfying assignment, then $p(\Phi)\in\ucol$ and hence
$\force{p(\Phi)}\le2$, while if $\Phi$ is unsatisfiable, then
$p(\Phi)\notin\cccol$ and hence $\force{p(\Phi)}>2$.

Valiant and Vazirani \cite{VVa} designed a polynomial-time computable
randomized transformation $r$ of the set of Boolean formulas such that,
if $\Phi$ is a satisfiable formula, then with a non-negligible probability
the formula $r(\Phi)$ has a unique satisfying assignment, while if $\Phi$ is
unsatisfiable, then $r(\Phi)$ is surely unsatisfiable.
Combining $r$ with the parsimonious reduction $p$ of \sat\/ to \cccol,
we arrive at the conclusion that \decforce{2} is NP-hard with respect
to {\em randomized\/} polynomial-time many-one reductions.
As a consequence, the forcing
chromatic number is not computable in polynomial time unless
any problem in NP is solvable by a polynomial-time Monte Carlo algorithm
with one-sided error.

We aim at determining the computational complexity of $\force G$
more precisely. Our first result establishes the hardness of
\decforce{2} with respect to {\em deterministic\/} polynomial-time
many-one reductions. The latter reducibility concept will be default
in what follows. The complexity class US, introduced by Blass and Gurevich
\cite{BGu}, consists of languages $L$ for which there is a
polynomial-time nondeterministic Turing machine $N$ such that
a word $x$ belongs to $L$ iff $N$ on input $x$ has exactly one
accepting computation path. Denote the set of Boolean formulas with
exactly one satisfying assignment by \usat. This set is complete for
US. As easily seen, \ucol\/ belongs to US and, by the parsimonious
reduction from \sat\/ to \cccol, \ucol\/ is another US-complete set.
By the Valiant-Vazirani reduction, the US-hardness under polynomial-time
many-one reductions implies the NP-hardness under randomized reductions
and hence the former hardness concept should be considered stronger.
It is known that US includes coNP \cite{BGu} and this inclusion is
proper unless the polynomial time hierarchy collapses \cite{RCR}.
Thus, the US-hardness implies also the coNP-hardness.
We prove that the problem \decforce{2} is US-hard.

Note that this result is equivalent to the reducibility of \ucol\/ to
\decforce2. Such a reduction would follow from the naive hypothesis,
which may be suggested by \refeq{eq:simple}, that a 3-chromatic $G$
is in \ucol\/ iff $\force G=2$. It should be stressed that the latter
is far from being true in view of Lemma \ref{lem:basic}.3 below.

On the other hand, we are able to estimate the complexity of each
\decforce{k} from above by putting this family of problems in
a complexity class which is a natural extension of US.
We show that,
for each $k\ge2$, the problem \decforce{k} is  reducible to a set in US
via a polynomial-time disjunctive truth-table reduction ({\em dtt-reduction\/}
for brevity, see Section \ref{s:back} for definitions). 
This improves on the straightforward inclusion of \decforce{k}
in $\sitwo$.\\
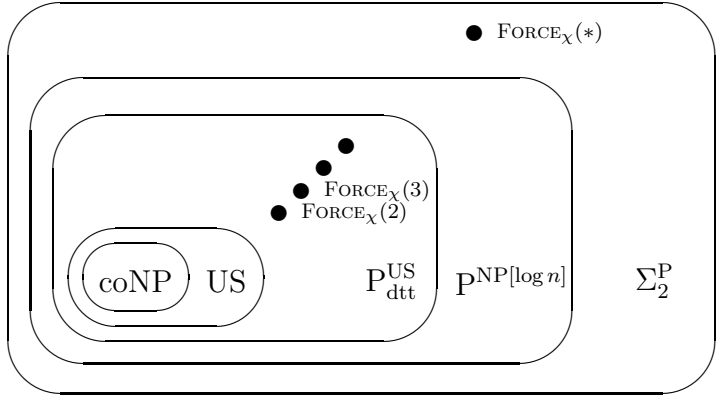
\begin{figure}
\centerline{
\unitlength=1.00mm
\special{em:linewidth 0.4pt}
\linethickness{0.4pt}
\begin{picture}(100.00,55.00)
\put(42.00,27.00){\circle*{2.00}}
\put(45.00,30.00){\circle*{2.00}}
\put(23.00,18.00){\makebox(0,0)[cc]{coNP}}
\put(45.00,27.00){\makebox(0,0)[lc]{{\scriptsize\decforce2}}}
\put(48.00,30.00){\makebox(0,0)[lc]{{\scriptsize\decforce{3}}}}
\put(23.00,18.50){\oval(14.00,9.00)[]}
\put(27.00,18.50){\oval(26.00,13.00)[]}
\put(35.00,18.00){\makebox(0,0)[cc]{US}}
\put(57.00,18.00){\makebox(0,0)[cc]{$\cloUS$}}
\put(73.00,18.00){\makebox(0,0)[cc]{$\pnplog$}}
\put(48.00,33.00){\circle*{2.00}}
\put(51.00,36.00){\circle*{2.00}}
\put(37.50,25.00){\oval(51.00,30.00)[]}
\put(45.00,26.00){\oval(72.00,38.00)[]}
\put(53.00,29.00){\oval(94.00,52.00)[]}
\put(92.00,18.00){\makebox(0,0)[cc]{$\Sigma_2^{\mbox{\scriptsize P}}$}}
\put(68.00,51.00){\circle*{2.00}}
\put(71.00,51.00){\makebox(0,0)[lc]{$\scriptsize\decforce{*}$}}
\end{picture}
}
\caption{Location of the slice decision problems for $\force G$
in the hierarchy of complexity classes.}
\end{figure}

Denote the class of decision problems reducible to  US under
dtt-reductions by $\cloUS$. As shown by
Chang, Kadin, and Rohatgi \cite{CKR}, $\cloUS$ is strictly larger than US
unless the polynomial time hierarchy collapses to its third level.
The position of the problems under consideration in the hierarchy of
complexity classes is shown in Figure~1, where $\pnplog$ denotes the class
of decision problems solvable by polynomial-time Turing machines
with logarithmic number of queries to an NP oracle. The latter class
coincides with the class of problems polynomial-time truth-table
reducible to NP, see \cite{Hem}.
\comm{That is, $\pnplog=\mbox{P}^{\mbox{\scriptsize NP}}_{tt}$.
This class has also name
$\mbox{P}^{\mbox{\scriptsize NP}}_{||}$, languages recognizable
in polynomial time with non-adaptive queries to an NP oracle.}
Another relation of $\cloUS$ to known complexity classes is
$\cloUS\subseteq\cpclass\subseteq\ppclass$, where 
a language $L$ is in \ppclass\ (resp.\ \cpclass)
if it is recognizable by a nondeterministic Turing machine $M$ with
the folowing acceptance criterion: an input word $w$ is in $L$
iff at least (resp.\ precisely) a half of the computing paths of $M$ on $w$
are accepting. This inclusion follows from the facts that $\US\subseteq\cpclass$
and that $\cpclass$ is closed under dtt-reductions
(see \cite[Theorem 9.9]{HOg}).

In a recent paper \cite{HMa}, Hatami and Maserrat obtain a result that,
in a sence, is complementary to our work and by this reason is also
shown in Figure 1. Let
$\decfor\chi{*}=\setdef{(x,k)}{F_\chi(x)\le k}$.
The authors of \cite{HMa} prove that the recognition of membership
in $\decfor\chi{*}$ is a $\sitwo$-complete problem.
Note that \cite{HMa} and our paper use different techniques and, moreover, 
the two approaches apparently cannot be used in place of one another.

Our next result gives a finer information about the hardness of \decforce2.
Note that, if $\chi(G)=2$, then $\force G$ is equal to the number
of connected components of $G$. It turns out that the knowledge that
$\chi(G)=3$ does not help in computing $\force G$. Moreover, it is
hard to recognize whether or not $\force G=2$ even if it is known
that $\force G\le3$. Stating these strengthenings, we relax our
hardness concept from the US-hardness to the coNP-hardness
(as already mentioned, the former implies the latter but the converse is
not true unless the polynomial time hierarchy collapses).
Thus, we prove that
the problem \decforce{2} is coNP-hard even under the promises that
$G\in\cccol$ and $\force G\le3$ (see Figure~2).
Note that the Valiant-Vazirani reduction implies no kind of a hardness
result for the promise version of \decforce{2}.\\
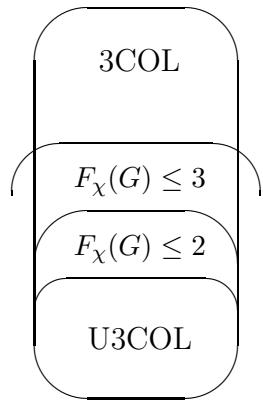
\begin{figure}
\centerline{
\unitlength=1.00mm
\special{em:linewidth 0.4pt}
\linethickness{0.4pt}
\begin{picture}(49.00,55.00)
\put(32.50,29.00){\oval(27.00,52.00)[]}
\put(32.50,14.50){\oval(27.00,9.00)[t]}
\put(32.50,20.50){\oval(27.00,15.00)[t]}
\put(33.00,11.00){\makebox(0,0)[cc]{\ucol}}
\put(33.00,23.00){\makebox(0,0)[cc]{{\small $\force G\le2$}}}
\put(33.00,32.00){\makebox(0,0)[cc]{{\small $\force G\le3$}}}
\put(33.00,48.00){\makebox(0,0)[cc]{\cccol}}
\put(32.50,30.00){\oval(33.00,14.00)[t]}
\end{picture}
}
\caption{The class $\setdef{G}{\force G\le2}$ and surroundings.}
\end{figure}

In fact, many other graph characteristics also have natural forcing
variants. Recall that a {\em clique\/} in a graph
is a set of pairwise adjacent vertices. The maximum number of vertices
of $G$ in a clique is denoted by $\om(G)$ and called the {\em clique
number\/} of $G$. A clique is {\em optimal\/}
if it consists of $\om(G)$ vertices. A set of vertices is called
{\em forcing\/} if it is included in a unique optimal clique.
We denote the minimum cardinality of a forcing set by
$F_\om(G)$ and call it the {\em forcing clique number\/} of~$G$.

Furthermore, we say that a vertex of a graph $G$ {\em dominates\/}
itself and any adjacent vertex. A set $D\subseteq V(G)$ is called
{\em dominating\/} if every vertex of $G$ is dominated by a vertex in $D$.
The {\em domination number\/} of $G$, denoted by $\gh(G)$, is the
minimum cardinality of a dominating set of $G$. Similarly to the above,
a {\em forcing set\/} of vertices is one included in a unique optimal
dominating set. The minimum cardinality of a forcing set is denoted by
$F_\gh(G)$ and called the {\em forcing domination number\/} of~$G$.
This graph invariant is introduced and studied by
Chartrand, Gavlas, Vandell, and Harary~\cite{CGVH}.

For the forcing clique and domination numbers we consider the
respective slice decision problems \decfor\om{k} and \decfor\gh{k}
and show the same relation of them to the class US that we have
for the forcing chromatic number. Actually, the
dtt-reducibility to US is proved for all the three numbers by a uniform
argument. However, the US-hardness with respect to many-one reductions
for $\om$ and $\gh$ is proved differently than for $\chi$. The case of $\om$ and $\gh$
seems combinatorially simpler because of the following equivalence:
A graph $G$ has a unique optimal clique iff $F_\om(G)=0$ and similarly
with $\gh$. The study of {\em unique optimum (UO) problems\/} was
initiated by Papadimitriou \cite{Pap}. Due to the US-hardness of
the UO CLIQUE and UO DOMINATING SET problems, we are able to show
the US-hardness of \decfor\om{k} and \decfor\gh{k} using only well-known
standard reductions, whereas for \decforce{k} we use somewhat more
elaborate reductions involving graph products.

\subsection*{Overview of previous related work}

\paragraph{Forcing chromatic number of particular graphs.}

Let $K_n$ (resp.\ $C_n$, $P_n$) denote the complete graph (resp.\ the cycle,
the path) on $n$ vertices. As a simple exercise, we have
$\force{C_{2m+1}}=m+1$. 
Mahmoodian, Naserasr, and Zaker
\cite{MNZ} compute the forcing chromatic number
of several Cartesian products:
$\force{C_{2m+1}\times K_2}=m+1$,
$\force{C_{m}\times K_n}=m(n-3)$ and
$\force{P_{m}\times K_n}=m(n-3)+2$ if $n\ge6$, and
$\force{K_{m}\times K_n}=m(n-m)$ if $n\ge m^2$.
Mahdian, Mahmoodian, Naserasr, and Harary
\cite{MMNH} determine a few missing values:
$\force{C_{m}\times K_3}=\lfloor m/2\rfloor+1$ and, if $m$ even,
$\force{C_{m}\times K_5}=2m$.
On the other hand, the asymptotics of $\force{K_n\times K_n}$ are unknown
(a problem having arisen in research on Latin squares, see below).
The best lower and upper bounds 
$\lfloor 4(n-2)/3\rfloor\le\force{K_n\times K_n}\le n^2/4$
are obtained, respectively, in \cite{HAF} for $n\ge8$ and \cite{CvR} for all $n$. 
Our results show that no general approach for
efficient computing the forcing chromatic number is possible
unless $\mbox{\rm NP}=\mbox{\rm P}$ (and even $\mbox{\rm US}=\mbox{\rm P}$).

\paragraph{Latin squares and the complexity of recognizing a forcing coloring.}

A {\em Latin square of order $n$\/} is an $n\times n$ matrix with
entries in $\{1,2,\ldots,n\}$ such that every row and column contains
all the $n$ numbers. In a {\em partial Latin square\/} some entries
may be empty and every number occurs in any row or column at most once.
A partial Latin square is called a {\em critical set\/} if it can be
completed to a Latin square in a unique way.
Colbourn, Colbourn, and Stinson \cite{CCS} proved that recognition if
a given partial Latin square $L$ is a critical set is coNP-hard.
Moreover, the problem is coNP-complete even if one extension of $L$
to a Latin square is known.
The result of \cite{CCS} is strengthened in~\cite{EGP}.

As it is observed in \cite{MNZ}, there is a natural one-to-one
correspondence between Latin squares of order $n$ and proper
$n$-colorings of the Cartesian square $K_n\times K_n$ which
matches critical sets and forcing colorings.\footnote{%
Such a correspondence is also well known for edge colorings of the
complete bipartite graph~$K_{n,n}$.}
It follows that it is
coNP-hard to recognize if a given partial coloring $p$ of a graph is forcing.
Moreover, even if this problem is restricted to graphs $K_n\times K_n$
and one extension of $p$ to a proper $n$-coloring is given, the problem
is coNP-complete (see also Proposition \ref{prop:fcrecogn} below).

\paragraph{Complexity of the forcing matching number}
Given a graph $G$ with a perfect matching, Harary, Klein, and \v{Z}ivkovi\'c
\cite{HKZ} define its forcing matching number as the minimum size of
a forcing set of edges, where the latter is a set
contained in a unique perfect matching.
Denote this number by $F_\nu(G)$. Let $\decfor{\nu}k$
be the problem of recognition, given $G$, if $F_\nu(G)\le k$
and let $\decfor{\nu}{*}$ be the variant of the same problem with
$k$ given as a part of an input. From the polynomial-time solvability
of the perfect matching problem, it easily follows that each $\decfor{\nu}k$
is polynomial-time solvable and that $\decfor{\nu}{*}$ is in NP.
Afshani, Hatami, and Mahmoodian \cite{AHM}, using an earlier result by
Adams, Mahdian, and Mahmoodian \cite{AMM}, prove that the latter problem
is NP-complete.

\paragraph{Variety of combinatorial forcing numbers.}

Critical sets are studied since the seventies (Nelder \cite{Nel}).
The forcing chromatic, domination, and matching numbers
attracted attention of researchers in the nineties. 
In fact, a number of other problems in diverse
areas of combinatorics have a similar forcing nature.
Defining sets in block designs (Gray \cite{Gra}) have a rather rich bibliography. 
Other graph invariants
whose forcing versions have appeared in the literature are
the orientation number (Chartrand, Harary, Schultz, and Wall \cite{CHSW}),
the geodetic number, the hull number, the dimension of a graph
(Chartrand and Zhang, respectively, \cite{CZh1,CZh2,CZh3}), and this list is still inexhaustive.
As a general concept, the combinatorial forcing was presented by
Harary in a series of talks, e.g.~\cite{Har}.

\paragraph{Unique colorability.}

The concept of a uniquely colorable graph was introduced by
Harary, Hedetniemi, and Robinson \cite{HHR}.
Complexity-theoretic concepts related to this combinatorial
phenomenon were introduced by Valiant and Vazirani \cite{VVa}
and Blass and Gurevich \cite{BGu}. However, the complexity-theorists
prefer to deal with \usat, the uniqueness version of the archetypical
\sat isfiability problem. The results established for \usat, as US- and
coNP-hardness under many-one reductions and NP-hardness under randomized
reductions, carry through for \ucol\/ by means of the parsimonious
many-one reduction of \sat\/ to \cccol. A direct, purely combinatorial
way to show the coNP-hardness of \ucol\/ is given by a result of
Greenwell and Lov\'asz (\cite{GLo}, see also \cite[Theorem 8.5]{IKl}).
They prove, in particular, that if a connected graph $G$ is not 3-colorable
then the categorical product $G\cdot K_3$ is a uniquely colorable graph.
On the other hand, if $G$ is 3-colorable then $G\cdot K_3$ can be colored
in two different ways. The latter follows from a simple observation:
any proper 3-coloring of one of the factors efficiently induces a
proper 3-coloring of the product $G_1\cdot G_2$. By the NP-completeness
of \cccol, we arrive at the conclusion that the problem of recognizing,
given a graph $H$ and its proper 3-coloring, whether or not $H$ is
uniquely 3-colorable, is coNP-complete. Later this complexity-theoretic
fact was observed by Dailey \cite{Dai} who uses an identical combinatorial argument.

\subsection*{Organization of the paper}

In Section \ref{s:back} we
define the categorical and the Cartesian graph products, which will play an
important role in our proofs, and prove some preliminary lemmas about
forcing sets in product graphs.
We also state a few basic bounds for $\force G$ and
determine the complexity of recognition if a given set of vertices
in a graph is forcing (the latter result is not used in the sequel
but is worth being noticed).
The hardness of \decforce{k} is established in Section \ref{s:forceklower}.
A closer look at \decforce{2} is taken in Section \ref{s:force2}.
In Section \ref{s:maxmin}, using a combinatorial result of
Hajiabolhassan, Mehrabadi, Tusserkani, and Zaker \cite{HMTZ},
we analyze the complexity of a related
graph invariant, namely, the largest cardinality of an inclusion-minimal
forcing set. Before taking into consideration the forcing clique
and domination numbers, we suggest a general setting for forcing
combinatorial numbers in Section \ref{s:set}. It is built upon
the standard formal concept of an NP optimization problem.
We benefit from this formal framework in some proofs.
Section \ref{s:omgg} is devoted to the forcing clique
and domination numbers. The dtt-reducibility of \decfor\pi{k} to US
for $\pi\in\{\chi,\om,\gh\}$ is shown in Section \ref{s:forcekupper}.
Section~\ref{s:open} contains a concluding discussion and some
open questions.

\section{Background}\label{s:back}

\subsection{Basics of complexity theory}

We suppose that the discrete structures under consideration are
encoded by binary words. For example, graphs are naturally representible
by their adjacency matrices. The set of all binary words is denoted by
$\{0,1\}^*$. A decision problem is identified with a language, i.e., a
subset of $\{0,1\}^*$, consisting of all yes-instances.
A {\em many-one reduction\/} of a problem $X$ to a problem $Y$
is a map $r\function{\{0,1\}^*}{\{0,1\}^*}$ such that $x\in X$
iff $r(x)\in Y$. If $r(x)$ is computable in time bounded by a polynomial
in the length of $x$, the reduction is called {\em polynomial-time}.
We write $X\mone Y$ to say that there is a polynomial-time many-one
reduction from $X$ to~$Y$.

Let $C$ be a class of decision problems (or languages). A problem $Z$
is called {\em $C$-hard\/} if any $X$ in $C$ is $\mone$-reducible to $Z$.
A problem $Z$ is called {\em $C$-complete\/} if $Z$ is
$C$-hard and belongs to $C$. \usat\/ and \ucol\/ are examples of
US-complete problems.

A {\em disjunctive truth-table reduction\/} (or {\em dtt-reduction\/}) of a language $X$
to a language $Y$ is a transformation which takes any word $x$
to a set of words $y_1,\ldots,y_m$ so that $x\in X$ iff
$y_i\in Y$ for at least one $i\le m$.
We write $X\disj Y$ to say that there is such a polynomial-time reduction
from $X$ to~$Y$.

If C is a class of languages and $\le$ is a reducibility, then
$\mbox{C}\le X$ means that $Y\le X$ for all $Y$ in C
(i.e., $X$ is C-hard under $\le$) and $X\le\mbox{C}$ means that
$X\le Y$ for some $Y$ in~C.

The formal framework of promise problems is developed in \cite{Sel}.
Let $Y$ and $Q$ be languages. Whenever referring to a {\em decision
problem $Y$ under the promise $Q$}, we mean that membership in $Y$
is to be decided only for inputs in $Q$. A reduction $r$ of an ordinary
decision problem $X$ to a problem $Y$ under the promise $Q$ is a usual
many-one reduction form $X$ to $Y$ with the additional requirement that
$r(x)\in Q$ for all $x$. This definition allows us to extend the
notion of $C$-hardness to promise problems.

A polynomial-time computable function
$h\function{\{0,1\}^*\times\{0,1\}^*}{\{0,1\}^*}$ is called an
{\em AND$_2$ function for a language $Z$\/} if for any pair $x,y$ we have
both $x$ and $y$ in $Z$ iff $h(x,y)$ is in $Z$. Such an $h$ is
an {\em OR$_2$ function for $Z$\/} if we have at least one of
$x$ and $y$ in $Z$ iff $h(x,y)$ is in~$Z$.

\subsection{Graph products}

Let $E(G)$ denote the set of edges of a graph $G$.
Given two graphs $G_1$ and $G_2$, we define a product graph on the vertex set
$V(G_1)\times V(G_2)$ in two ways. Vertices
$(u_1,u_2)$ and $(v_1,v_2)$ are adjacent in the {\em Cartesian
product\/} $G_1\times G_2$ if either $u_1=v_1$ and $\{u_2,v_2\}\in E(G_2)$
or $u_2=v_2$ and $\{u_1,v_1\}\in E(G_1)$. They are adjacent
in the {\em categorical product\/} $G_1\cdot G_2$ if both
$\{u_1,v_1\}\in E(G_1)$ and $\{u_2,v_2\}\in E(G_2)$.

A set $V(G_1)\times\{v\}$ for $v\in V(G_2)$ will be called
a {\em $G_1$-layer of $v$\/} and a set $\{u\}\times V(G_2)$ for
$u\in V(G_1)$ will be called a {\em $G_2$-layer of $u$}.

\begin{lemma}\label{lem:chiofchart}
{\bf (Sabidussi, see \cite[Theorem 8.1]{IKl})}
$\chi(G\times H)=\max\{\chi(G),\chi(H)\}.$
\end{lemma}

If $c$ is a proper coloring of $G$, it is easy to see that $c^*(x,y)=c(x)$
is a proper coloring of the categorical product $G\cdot H$. 
We will say that $c$ {\em induces\/}
$c^*$. Similarly, any proper coloring of $H$ induces a proper coloring
of $G\cdot H$. This implies the following well-known fact.

\begin{lemma}\label{lem:chiofcat}
$\chi(G\cdot H)\le\min\{\chi(G),\chi(H)\}.$
\end{lemma}

The next proposition shows that the Cartesian and the categorical
products are, respectively, AND$_2$ and OR$_2$ functions for \cccol\/
(see \cite{KST} for an exposition of AND and OR functions).

\begin{proposition}
\begin{venumerate}
\item
$G\times H\in\cccol$ iff both $G$ and $H$ are in \cccol.
\item
$G\cdot H\in\cccol$ iff at least one of $G$ and $H$ is in \cccol.
\end{venumerate}
\end{proposition}

\begin{proof}
Item 1 is straightforward by Lemma \ref{lem:chiofchart}.
However, Item 2 does not follow solely from Lemma \ref{lem:chiofcat}
because the equality $\chi(G\cdot H)=\min\{\chi(G),\chi(H)\}$
is still an unproven conjecture made by Hedetniemi. Luckily, the conjecture
is known to be true for 4-chromatic graphs (El-Zahar and Sauer,
see \cite[Section 8.2]{IKl}), and this suffices for our claim.
\end{proof}

\subsection{Preliminary lemmas}

\begin{lemma}\label{lem:grlo}
{\bf (Greenwell-Lov\'asz \cite{GLo})}
Let $G$ be a connected graph with $\chi(G)>n$. Then
$G\cdot K_n$ is uniquely $n$-colorable.
\end{lemma}

\noindent
The proof can be found also in \cite[Theorem 8.5]{IKl}.
We will use not only Lemma \ref{lem:grlo} itself but also a component
of its proof stated as the next lemma.
We call a partial coloring {\em injective\/} if 
different colored vertices receive different colors.

\begin{lemma}\label{lem:c1}
Let $G$ be a connected graph and $p$ be an injective coloring of a
$K_n$-layer of $G\cdot K_n$.
Then $p$ forces the $K_n$-induced coloring of $G\cdot K_n$.
\end{lemma}

\begin{proof}
Suppose that $p$ is an injective coloring of a $K_n$-layer of $u\in V(G)$
and that a proper $n$-coloring $c$ of $G\cdot K_n$ extends $p$.
Let $v\in V(G)$ be adjacent to $u$. Inferably, $c(v,i)=c(u,i)$
for any $i\in V(K_n)$. Since $G$ is connected,
$c$ is forced by~$p$ to be monochrome on each $G$-layer.
\end{proof}

\begin{lemma}\label{lem:c2}
Any proper 3-coloring of $K_3\cdot K_3$ is induced by one of
the two factors~$K_3$.
\end{lemma}

\begin{proof}
Let $V(K_3)=\{1,2,3\}$ and denote $L_i=\{(i,1),(i,2),(i,3)\}$.
Suppose that $c$ is a coloring of $K_3\cdot K_3$. Consider it on one of the
layers, say, $L_2$. If $|c(L_2)|=3$, then $c$ is
induced by the second factor on the account of Lemma \ref{lem:c1}.
Assume that $|c(L_2)|=2$, for example, $c(2,1)=1$, $c(2,2)=2$, and
$c(2,3)=2$. Then $c(1,2)=c(3,3)=3$ is forced, contradictory to the fact
that these vertices are adjacent. There remains the possibility that
$|c(L_2)|=1$, for example, $c(2,1)=c(2,2)=c(2,3)=2$. Color 2 can
occur neither in $L_1$ nor in $L_3$. Assume that one of these
layers has both colors 1 and 3, say, $c(1,1)=1$ and $c(1,2)=3$.
This forces $c(3,3)=2$, a contradiction. Thus, $|c(L)|=1$ is possible
only if $L_1$, $L_2$, and $L_3$ are all monochrome and have
pairwise distinct colors. This is the coloring induced by the first
factor.
\end{proof}

\subsection{A few basic bounds}

We call two $s$-colorings {\em equivalent\/} if they are obtainable from
one another by permutation of colors. Proper $s$-colorings of a graph $G$
are equivalent if they determine the same partition of $V(G)$ into
$s$ independent sets. Let $\numcol G$ denote the number of such partitions
for $s=\chi(G)$. Thus, $\numcol G$ is equal to the number of inequivalent
proper $s$-colorings of $s$-chromatic $G$, while the total number of such
colorings is equal to $\chi(G)!\numcol G$.
A graph $G$ is {\em uniquely colorable\/} if $\numcol G=1$.
In particular, $G\in\ucol$ iff $\chi(G)=3$ and $\numcol G=1$.

\begin{lemma}\label{lem:basic}
\mbox{}

\begin{venumerate}
\item
$\chi(G)-1\le\force G\le\log_2\numcol G+\log_2(\chi(G)!)$.
\item
If $\numcol G=1$, then $\force G=\chi(G)-1$.
\item
For any $k$, there is a 3-chromatic graph $G_k$ on $4k+2$ vertices 
with $\force{G_k}=2$ and $\numcol{G_k}=2^{k-1}+1$.
\end{venumerate}
\end{lemma}

\noindent
The lower bound in Item 1 is sharp by Item 2. The upper bound is sharp
because, for example, $\force{mK_2}=m$ while $\numcol{mK_2}=2^{m-1}$, where
$mK_2$ denotes the graph consisting of $m$ isolated edges.
Item 3 shows that the converse of Item 2 is false.
It is also worth noting that both the lower and the upper bounds in
Item 1 are computationally hard, even if one is content with finding an
approximate value. The NP-hardness of approximation of the chromatic number
is established in \cite{LYa}.
\comm{
(see also \cite{FKi} for the best of subsequent results). 
}
A hardness result for approximate computing
$\log_2\numcol G$, even for 3-chromatic graphs, is obtained in~\cite{Zuc}.

\begin{proof}
Item 2 and the lower bound in Item 1 are obvious. To prove the upper
bound in Item 1, we have to show that a $k$-chromatic graph $G$ has
a forcing set  of at most $l=\lfloor\log_2(k!\numcol G)\rfloor$ vertices
$v_1,v_2,\ldots$. Let $C_1$ be the set
of all $k!\numcol G$ proper $k$-colorings of $G$. We choose vertices
$v_1,v_2,\ldots$ one by one as follows. Let $i\ge1$. Assume that the
preceding $i-1$ vertices have been chosen and that a set of colorings
$C_i$ has been defined and has at least 2 elements. We set $v_i$ to be
an arbitrary vertex such that not all colorings in $C_i$ coincide at $v_i$.
Furthermore, we assign $v_i$ a color $p(v_i)$ occurring in the multiset
$\setdef{c(v_i)}{c\in C_i}$ least frequently. Finally, we define
$C_{i+1}=\setdef{c\in C_i}{c(v_i)=p(v_i)}$. Note that
$|C_{i+1}|\le|C_i|/2$. We eventually have $|C_{i+1}|=1$ for some
$i\le l=\lfloor\log_2|C_1|\rfloor$. For this $i$, denote the single
coloring in $C_{i+1}$ by $c$. By construction, $v_1,\ldots,v_i$ is
defining for~$c$.

To prove Item 3, consider $H=K_3\times K_2$. This graph has two
inequivalent colorings $c_1$ and $c_2$ shown in Figure 3.
Let $u,v,w\in V(H)$ be as in Figure 3. Note that a partial coloring
$p_1(u)\ne p_1(v)$ forces $c_1$ or its equivalent and that
$p_2(u)=p_2(v)\ne p_2(w)$ forces $c_2$.

Let $G_k$ consist of $k$ copies of $H$ with all $u$ and all $v$
identified, that is, $G_k$ has $4k+2$ vertices. Since the set
$\{u,v\}$ stays forcing in $G_k$, we have $\force{G_k}=2$. If
$u$ and $v$ are assigned the same color, we are free to assign
each copy of $w$ any of the two remaining colors. It follows that
$\numcol{G_k}=2^{k-1}+1$.
\end{proof}

\noindent
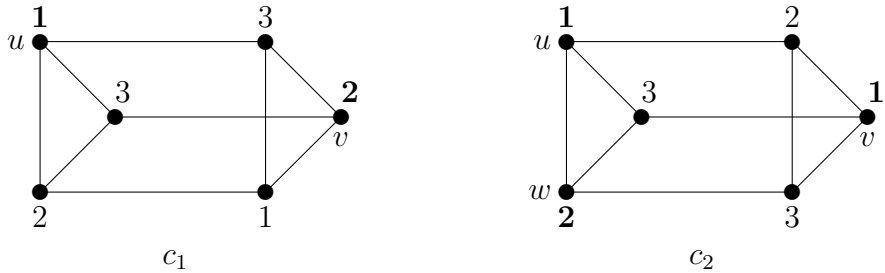
\begin{figure}
\centerline{
\unitlength=1.00mm
\special{em:linewidth 0.4pt}
\linethickness{0.4pt}
\begin{picture}(121.00,32.00)
\emline{10.00}{10.00}{1}{20.00}{20.00}{2}
\emline{20.00}{20.00}{3}{10.00}{30.00}{4}
\emline{10.00}{30.00}{5}{10.00}{10.00}{6}
\put(10.00,10.00){\circle*{2.00}}
\put(20.00,20.00){\circle*{2.00}}
\put(10.00,30.00){\circle*{2.00}}
\emline{40.00}{10.00}{7}{50.00}{20.00}{8}
\emline{50.00}{20.00}{9}{40.00}{30.00}{10}
\emline{40.00}{30.00}{11}{40.00}{10.00}{12}
\put(40.00,10.00){\circle*{2.00}}
\put(50.00,20.00){\circle*{2.00}}
\put(40.00,30.00){\circle*{2.00}}
\emline{40.00}{10.00}{13}{10.00}{10.00}{14}
\emline{10.00}{30.00}{15}{40.00}{30.00}{16}
\emline{20.00}{20.00}{17}{50.00}{20.00}{18}
\emline{80.00}{10.00}{19}{90.00}{20.00}{20}
\emline{90.00}{20.00}{21}{80.00}{30.00}{22}
\emline{80.00}{30.00}{23}{80.00}{10.00}{24}
\put(80.00,10.00){\circle*{2.00}}
\put(90.00,20.00){\circle*{2.00}}
\put(80.00,30.00){\circle*{2.00}}
\emline{110.00}{10.00}{25}{120.00}{20.00}{26}
\emline{120.00}{20.00}{27}{110.00}{30.00}{28}
\emline{110.00}{30.00}{29}{110.00}{10.00}{30}
\put(110.00,10.00){\circle*{2.00}}
\put(120.00,20.00){\circle*{2.00}}
\put(110.00,30.00){\circle*{2.00}}
\emline{110.00}{10.00}{31}{80.00}{10.00}{32}
\emline{80.00}{30.00}{33}{110.00}{30.00}{34}
\emline{90.00}{20.00}{35}{120.00}{20.00}{36}
\put(10.00,32.00){\makebox(0,0)[cb]{{\bf 1}}}
\put(10.00,8.00){\makebox(0,0)[ct]{2}}
\put(20.00,22.00){\makebox(0,0)[lb]{3}}
\put(8.00,30.00){\makebox(0,0)[rc]{$u$}}
\put(40.00,32.00){\makebox(0,0)[cb]{3}}
\put(40.00,8.00){\makebox(0,0)[ct]{1}}
\put(50.00,22.00){\makebox(0,0)[lb]{{\bf 2}}}
\put(50.00,18.00){\makebox(0,0)[ct]{$v$}}
\put(80.00,32.00){\makebox(0,0)[cb]{{\bf 1}}}
\put(80.00,8.00){\makebox(0,0)[ct]{{\bf 2}}}
\put(90.00,22.00){\makebox(0,0)[lb]{3}}
\put(110.00,32.00){\makebox(0,0)[cb]{2}}
\put(110.00,8.00){\makebox(0,0)[ct]{3}}
\put(120.00,22.00){\makebox(0,0)[lb]{{\bf 1}}}
\put(28.00,1.00){\makebox(0,0)[cc]{$c_1$}}
\put(98.00,1.00){\makebox(0,0)[cc]{$c_2$}}
\put(78.00,30.00){\makebox(0,0)[rc]{$u$}}
\put(120.00,18.00){\makebox(0,0)[ct]{$v$}}
\put(78.00,10.00){\makebox(0,0)[rc]{$w$}}
\end{picture}
}
\caption{Proper 3-colorings of $K_3\times K_2$.}
\end{figure}

\subsection{Complexity of a forcing set recognition}

\begin{proposition}\label{prop:fcrecogn}
The problem of recognizing, given a graph $G$ and two vertices $u,v\in V(G)$,
whether or not $\{u,v\}$ is a forcing set is US-complete.
\end{proposition}

\begin{proof}
Suppose that $\chi(G)\ge3$ for else the question is efficiently
decidable. Set $p(u)=1$ and $p(v)=2$. Obviously,
$\{u,v\}$ is a forcing set iff $p$ forces a proper 3-coloring of $G$.
Verification of the latter condition is clearly in~US.

We now describe a reduction $R$ of the US-complete problem \ucol\/
to the problem under consideration. If an input graph $G$ is empty,
let $R(G)$ be an arbitrary graph with two vertices which are not
a forcing set. Otherwise, let $u$ and $v$ be the lexicographically first
pair of adjacent vertices in $G$. We set $R(G)=(G,u,v)$. It is not hard
to see that $G\in\ucol$ iff $R(G)$ consists of a graph and a 2-vertex
forcing set in it.
\end{proof}

\section{Complexity of $\force G$: A lower bound}\label{s:forceklower}

\begin{theorem}\label{thm:f1lower}
For each $k\ge2$, the problem \decforce{k} is US-hard.
Moreover, this holds true even if we consider only connected graphs.
\end{theorem}

We first observe that the family of problems
\decforce{k} is linearly ordered with respect to the $\mone$-reducibility.
A simple reduction showing this does not
preserve connectedness of graphs. However, if we restrict ourselves
to connected graphs, we are able to show that \decforce{2} remains
the minimum element in this order. We then
prove that \decforce{2} is US-hard (even for connected graphs).

\begin{lemma}\label{lem:red1}
$\decforce{k}\mone\decforce{k+1}$.
\end{lemma}

\begin{proof}
Given a non-empty graph $G$, we add one isolated vertex to it.
Denoting the result by $G+K_1$, it is enough to notice that
$\force{G+K_1}=\force G +1$.
\end{proof}

\begin{lemma}\label{lem:red2}
Let $k\ge2$. Then \decforce{2} reduces to \decforce{k} even if we consider
the problems only for connected graphs.
\end{lemma}

\begin{proof}
Let $G$ be a graph on $n$ vertices and $m\le n$. Writing $H=G\oplus mK_2$,
we mean that
\begin{itemize}
\item
$V(H)=\{v_1,\ldots,v_n\}\cup\bigcup_{i=1}^m\{a_i,b_i\}$,
\item
$H$ induces on $\{v_1,\ldots,v_n\}$ a graph isomorphic to $G$,
\item
$\{v_i,a_i\}$ and $\{a_i,b_i\}$ for all $i\le m$ are edges of $H$, and
\item
$H$ has no other edges.
\end{itemize}
Suppose that $\chi(G)\ge3$ and $H=G\oplus mK_2$. Let us check that
$\force H\le 2+m$ if $\force G=2$ and
$\force H\ge3+m$ if $\force G\ge3$. This will give us the following
reduction of \decforce{2} to \decforce{k}: Given $G$, construct an
$H=G\oplus(k-2)K_2$.

Let $\force G=2$. Note that $G$ must be 3-chromatic. To show that
$\force H\le 2+m$, we construct a forcing set of $2+m$ vertices.
We will identify $G$ and its copy spanned by $\{v_1,\ldots,v_n\}$ in $H$.
Let $v_j$ and $v_l$ be two vertices such that a partial coloring
$p(v_j)=1$ and $p(v_l)=2$ forces a proper 3-coloring $c$ of $G$.
Color each $b_i$ differently from $c(v_i)$. Then $\{v_j,v_l,b_i,\ldots,b_m\}$
becomes a forcing set in~$H$.

Let $\force G\ge3$. Clearly, any forcing set contains all $b_i$'s.
We hence have to show that no set
$S=\{b_1,\ldots,b_m,x,y\}\subset V(H)$ is forcing in $H$.
We will assume that $\chi(G)=3$ for else the claim is easy.
Suppose that $x\in\{v_j,a_j\}$ and $y\in\{v_l,a_l\}$ for some $j,l$ 
and that $S$ is augmented
with a coloring $p$ extendable to a proper
3-coloring $c$ of $H$. Let $c'$ denote the restriction of $c$ to
$\{v_1,\ldots,v_n\}$. Since $c'$ is not a unique coloring of $G$
compatible with $p$ (otherwise $\{v_j,v_l\}$ would be a forcing set for $G$), 
there is another such coloring $c''$.
Obviously, $c''$ and $p$ have a common extension to a proper
coloring of $H$. Thus, $S$ is not a forcing set.
\end{proof}

\comm{
Lemmas \ref{lem:red1} and \ref{lem:red2} hold true if we redefine
\decforce{k} as the problem of recognizing whether or not $\force G=k$.
In the proof of the latter lemma it is enough to notice that
$\force H=2+m$ if $\force G=2$.
To show that $\force H>1+m$, we have to prove that no $(1+m)$-vertex set
$S\subset V(H)$ can be forcing. Clearly, any forcing set contains all
$b_i$'s. Suppose that the remaining element of $S$ is $v_l$ or $a_l$
and that $S$ is augmented with a coloring $p$ extendable to a proper
3-coloring $c$ of $H$. Let $c'$ denote the restriction of $c$ to
$\{v_1,\ldots,v_n\}$ and let $c''$ be the result of interchanging
in $c'$ the two colors different from $c(v_l)$. As easily seen, $c''$
and $p$ agree and have a common extension to a proper 3-coloring of $H$.
Thus, $p$ has two proper extensions and hence is not forcing.
}

\begin{lemma}\label{lem:force2}
\decforce{2} is US-hard even if restricted to connected graphs.
\end{lemma}

To prove the lemma, we describe a reduction from \ucol.
Note that \ucol\/ remains US-complete when restricted to connected graphs
and that our reduction will preserve connectedness.
Since the class of 2-colorable graphs is tractable and can be excluded
from consideration, the desired reduction is given by the following lemma.

\begin{lemma}\label{lem:proof1}
Suppose that $\chi(G)\ge3$. Then $G\in\ucol$ iff $\force{G\times K_3}=2$.
\end{lemma}

\noindent
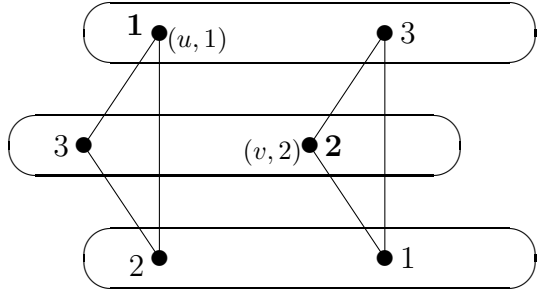
\begin{figure}
\centerline{
\unitlength=1.00mm
\special{em:linewidth 0.4pt}
\linethickness{0.4pt}
\begin{picture}(73.00,41.00)
\emline{13.00}{22.00}{1}{23.00}{37.00}{2}
\emline{23.00}{37.00}{3}{23.00}{7.00}{4}
\emline{23.00}{7.00}{5}{13.00}{22.00}{6}
\put(13.00,22.00){\circle*{2.00}}
\put(23.00,37.00){\circle*{2.00}}
\put(23.00,7.00){\circle*{2.00}}
\emline{43.00}{22.00}{7}{53.00}{37.00}{8}
\emline{53.00}{37.00}{9}{53.00}{7.00}{10}
\emline{53.00}{7.00}{11}{43.00}{22.00}{12}
\put(43.00,22.00){\circle*{2.00}}
\put(53.00,37.00){\circle*{2.00}}
\put(53.00,7.00){\circle*{2.00}}
\put(43.00,37.00){\oval(60.00,8.00)[]}
\put(33.00,22.00){\oval(60.00,8.00)[]}
\put(43.00,7.00){\oval(60.00,8.00)[]}
\put(24.00,36.00){\makebox(0,0)[lc]{{\footnotesize $(u,1)$}}}
\put(42.00,21.00){\makebox(0,0)[rc]{{\footnotesize $(v,2)$}}}
\put(21.00,38.00){\makebox(0,0)[rc]{{\bf 1}}}
\put(11.00,22.00){\makebox(0,0)[rc]{3}}
\put(21.00,6.00){\makebox(0,0)[rc]{2}}
\put(55.00,37.00){\makebox(0,0)[lc]{3}}
\put(55.00,7.00){\makebox(0,0)[lc]{1}}
\put(45.00,22.00){\makebox(0,0)[lc]{{\bf 2}}}
\end{picture}
}
\caption{Proof of Lemma \protect\ref{lem:proof1} (Case 1).}
\end{figure}

\begin{proof}
\case 1 {$G\in\ucol$}
We have to show that $\force{G\times K_3}=2$.

Fix arbitrary $u,v\in V(G)$ whose colors in the proper 3-coloring of $G$
are different, for example, $u$ and $v$ can be any adjacent vertices of $G$.
Let $V(K_3)=\{1,2,3\}$. Assign $p(u,1)=1$ and $p(v,2)=2$ and check that
$p$ forces a proper 3-coloring of $G\times K_3$. Assume that $c$ is
a proper 3-coloring of $G\times K_3$ consistent with $p$. Since
$c$ on each $G$-layer coincides with the 3-coloring of $G$ up to
permutation of colors, we easily infer that $c(v,1)=c(u,2)=3$
(see Figure 4). This implies $c(u,3)=2$ and $c(v,3)=1$. Thus,
in each $G$-layer we have two vertices with distinct colors, which determines
colors of all the other vertices. As easily seen, the coloring obtained
is really proper.

\case 2 {$G\in\cccol\setminus\ucol$}
We have to check that $\force{G\times K_3}\ge3$.

Given a partial coloring $p$ of two vertices $a,b\in V(G\times K_3)$,
we have to show that it is not forcing. The cases that $p(a)=p(b)$
or that $a$ and $b$ are in the same $G$- or $K_3$-layer are easy.
Without loss of generality we therefore suppose that $p(a)=1$, $p(b)=2$,
$a=(u,1)$, and $b=(v,2)$, where $u$ and $v$ are distinct vertices of $G$.
Define two partial colorings of $G$ by $c_1(u)=c_1(v)=1$ and by
$c_2(u)=1$, $c_2(v)=3$.

\subcase{2.1}{Both $c_1$ and $c_2$ extend to proper 3-colorings of~$G$}
Denote the extensions by $e_1$ and $e_2$ respectively. Denote the three
$G$-layers of $G\times K_3$ by $G_1,G_2,G_3$ and consider $e_1,e_2$
on $G_1$. For each $i=1,2$, $e_i$ and $p$ agree and have a common
extension to a proper coloring of $G\times K_3$. Thus, $p$ is not
forcing.

\subcase{2.2}{Only $c_1$ extends to a proper 3-coloring of~$G$}
Since $G$ is not uniquely colorable, there must be at least two
extensions, $e_1$ and $e_2$, of $c_1$ to proper 3-colorings of $G_1$.
As in the preceding case, $e_1$ and $e_2$ each agree with $p$
and together with $p$ extend two distinct colorings of $G\times K_3$.

\subcase{2.3}{Only $c_2$ extends to a proper coloring of~$G$}
This case is completely similar to Subcase~2.2.

\case 3 {$G\notin\cccol$}
We have $\chi(G\times K_3)\ge4$ by Lemma \ref{lem:chiofchart}
and $\force{G\times K_3}\ge3$ by Lemma~\ref{lem:basic}.1.
\end{proof}

Theorem \ref{thm:f1lower} immediately follows from Lemmas \ref{lem:force2}
and~\ref{lem:red2}.

\bigskip
\bigskip

\section{Hardness of \decforce{2}: A closer look}\label{s:force2}

\begin{theorem}\label{thm:f2}
The problem \decforce{2} is coNP-hard even under the promises that
$\force G\le3$ and $\chi(G)\le3$ and even if an input graph $G$
is given together with its proper 3-coloring.
\end{theorem}

Let us for a while omit the promise that $\force G\le3$.
Then the theorem is provable by combining
the Greenwell-Lov\'asz reduction of coNP to US (Lemma \ref{lem:grlo})
and our reduction of US to \decforce{2} (Lemma \ref{lem:proof1}).
Doing so, we easily deduce the following:
\begin{itemize}
\item
If $\chi(G)>3$, then $G\cdot K_3$ is uniquely 3-colorable and
hence $\force{(G\cdot K_3)\times K_3}=2$.
\item
If $\chi(G)=3$, then $G\cdot K_3$ is 3-chromatic
because it contains an odd cycle (this is an easy particular case
of the aforementioned Hedetniemi conjecture). Moreover, $G\cdot K_3$ 
has two induced 3-colorings and
hence $\force{(G\cdot K_3)\times K_3}\ge3$.
\end{itemize}
To obtain Theorem \ref{thm:f2} (without the promise
$\force G\le3$), it now suffices to make the following observation.

\begin{lemma}\label{lem:b}
$\chi((G\cdot K_3)\times K_3)=3$ for any graph $G$. Moreover, a proper
3-coloring is efficiently obtainable from the explicit product representation
of $(G\cdot K_3)\times K_3$.
\end{lemma}

\begin{proof}
By Lemma \ref{lem:chiofcat}, we have $\chi(G\cdot K_3)\le3$ and hence,
by Lemma \ref{lem:chiofchart}, $\chi((G\cdot K_3)\times K_3)=3$.
Let $V(K_3)=\{1,2,3\}$ and denote $V_{i,j}=\setdef{(v,i,j)}{v\in V(G)}$.
It is not hard to see that $V_{1,1}\cup V_{2,2}\cup V_{3,3}$,
$V_{1,2}\cup V_{2,3}\cup V_{3,1}$, and $V_{1,3}\cup V_{2,1}\cup V_{3,2}$
is a partition of $V((G\cdot K_3)\times K_3)$ into independent sets.
\end{proof}

\begin{remark}\rm
The known facts about graph factorizations
\cite[Chapters 4 and 5]{IKl} imply a nontrivial strengthening
of Lemma \ref{lem:b} under certain, rather general conditions. Namely,
if $G$ is a connected nonbipartite graph and no two vertices
of $G$ have the same neighborhood, then we do not need to
assume that $H=(G\cdot K_3)\times K_3$ is explicitly represented
as a product graph because the product stucture is efficiently
reconstructible from the isomorphism type of $H$.
We thank Wilfried Imrich for this observation.
\end{remark}

To obtain the full version of Theorem \ref{thm:f2}, we only slightly
modify the reduction: Before transforming $G$ in $(G\cdot K_3)\times K_3$,
we add to $G$ a triangle with one vertex in $V(G)$ and two vertices new. 
Provided $\chi(G)\ge3$, this does not change $\chi(G)$ and hence
the modified transformation is an equally good reduction. The strengthening
(the promise $\force G\le3$) is given by the following lemma.

\begin{lemma}\label{lem:proof2}
If a graph $G$ is connected and contains a triangle, then
$\force{(G\cdot K_3)\times K_3}\le3$.
\end{lemma}

\begin{proof}
Let $v$ be a vertex of a triangle $T$ in $G$. Consider the product
$H=G\cdot K_3$ and a partial coloring $p(v,1)=1$, $p(v,2)=2$.
We claim that $p$ forces the $K_3$-induced coloring of $H$.
Obviously, the latter is an extension of $p$. To show that no other
extension is possible, assume that $c$ is a proper 3-coloring of $H$
compatible with $p$ and consider the restriction of $c$ on $T\cdot K_3$.
By Lemma \ref{lem:c2}, $c$ on $T\cdot K_3$ coincides with the
coloring induced by the second factor. In particular, $c(v,3)=3$.
Our claim now follows from the connectedness of $G$ by Lemma~\ref{lem:c1}.\\
\begin{figure}
\centerline{
\unitlength=1.00mm
\special{em:linewidth 0.4pt}
\linethickness{0.4pt}
\begin{picture}(97.00,80.00)
\put(20.00,54.00){\oval(30.00,8.00)[]}
\put(20.00,64.00){\oval(30.00,8.00)[]}
\put(20.00,74.00){\oval(30.00,8.00)[]}
\put(80.00,54.00){\oval(30.00,8.00)[]}
\put(80.00,64.00){\oval(30.00,8.00)[]}
\put(80.00,74.00){\oval(30.00,8.00)[]}
\put(50.00,9.00){\oval(30.00,8.00)[]}
\put(50.00,19.00){\oval(30.00,8.00)[]}
\put(50.00,29.00){\oval(30.00,8.00)[]}
\put(15.00,74.00){\circle*{2.00}}
\put(15.00,64.00){\circle*{2.00}}
\put(81.00,74.00){\circle*{2.00}}
\put(30.00,54.00){\circle*{2.00}}
\put(90.00,54.00){\circle*{2.00}}
\put(60.00,9.00){\circle*{2.00}}
\emline{60.00}{9.00}{1}{30.00}{54.00}{2}
\emline{30.00}{54.00}{3}{90.00}{54.00}{4}
\emline{90.00}{54.00}{5}{60.00}{9.00}{6}
\put(13.00,74.00){\makebox(0,0)[rc]{{\bf 1}}}
\put(13.00,64.00){\makebox(0,0)[rc]{{\bf 2}}}
\put(16.00,73.00){\makebox(0,0)[lc]{{\footnotesize $(v,1,1)$}}}
\put(16.00,63.00){\makebox(0,0)[lc]{{\footnotesize $(v,2,1)$}}}
\put(79.00,74.00){\makebox(0,0)[rc]{{\bf 3}}}
\put(83.00,73.00){\makebox(0,0)[lc]{{\footnotesize $x$}}}
\put(4.00,80.00){\makebox(0,0)[rt]{$H_1$}}
\put(97.00,79.00){\makebox(0,0)[lt]{$H_2$}}
\put(67.00,3.00){\makebox(0,0)[lb]{$H_3$}}
\end{picture}
}
\caption{$(G\cdot K_3)\times K_3$ (Proof of Lemma \protect\ref{lem:proof2})}
\end{figure}
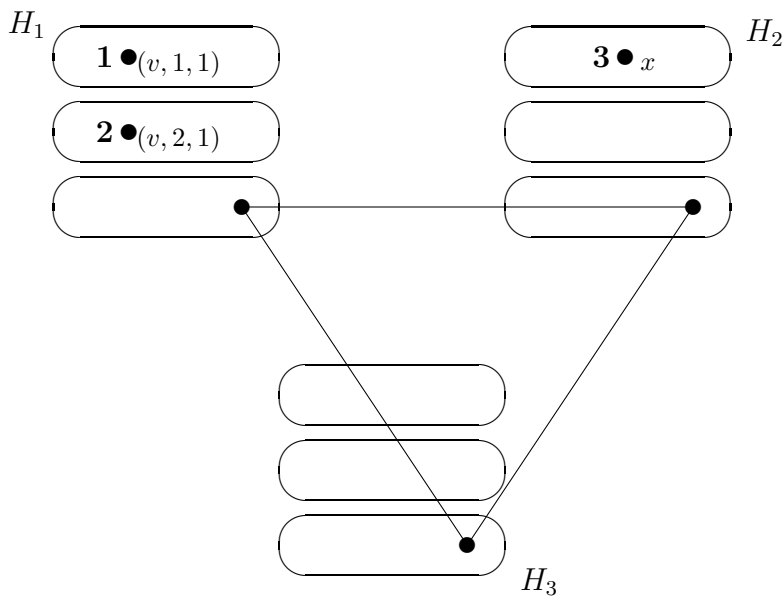

Now, let $H_1,H_2,H_3$ denote the three $H$-layers in $H\times K_3$ and,
for each $i=1,2,3$, let $G_{i1},G_{i2},G_{i3}$ denote the three $G$-layers
in $H_i$. Let $p(v,1,1)=1$, $p(v,2,1)=2$ be the forcing partial coloring
for $H_1$ as described above (see Figure 5). Thus, $p$ forces
coloring the whole $G_{1j}$ in color $j$ for each $j=1,2,3$.
Suppose that $c$ is a proper
3-coloring of $H\times K_3$ that agrees with $p$. From the product
structure of $H\times K_3$ we see that, for each $j$, in $G_{2j}$
there cannot occur color $j$. Let $T^2$ denote the copy of $T\cdot K_3$
in $H_2$. By Lemma \ref{lem:c2}, $c$ on $T^2$ is induced by the second
factor and hence
$c(v,1,2)$, $c(v,2,2)$, and $c(v,3,2)$ are pairwise distinct.
By Lemma \ref{lem:c1}, each of $G_{21}$, $G_{22}$, and $G_{23}$ is
monochrome and these layers receive pairwise distinct colors.
We already know that $c(G_{2j})\ne j$. Thus, when we define $p$ in the third
point by $p(x)=3$ for an arbitrary $x\in G_{21}$, this forces
$c(G_{21})=3$, $c(G_{22})=1$, and $c(G_{23})=2$. Since every vertex in
$G_{3j}$ is in triangle with its clones in $G_{1j}$ and $G_{2j}$,
$c$ is uniquely extrapolated on~$H_3$.
\end{proof}

The proof of Theorem \ref{thm:f2} is complete.

\section{Maximum size of a minimal forcing set}\label{s:maxmin}

Another related invariant of a graph $G$ is the largest cardinality
of an inclusion-minimal forcing set in $G$. We will denote this number
by $\fforce G$. A complexity analysis of $\fforce G$ is easier
owing to the characterization of uniquely colorable graphs obtained
in~\cite{HMTZ}.

\begin{lemma}\label{lem:haj}
{\bf (Hajiabolhassan, Mehrabadi, Tusserkani, and Zaker \cite{HMTZ})}
A connected graph $G$ is uniquely 3-colorable iff $\fforce G=2$.
\end{lemma}

\begin{theorem}\label{thm:f3}
The problem of deciding, given a graph $G$ and its proper 3-coloring,
whether or not $\fforce G\le 2$ is coNP-complete.
\end{theorem}

\begin{proof}
The problem is in coNP because a no-instance of it has a certificate
consisting of a proper 3-coloring $c$ of $G$ and a 3-vertex set
$A\subset V(G)$ such that no 2-element subset $B$ of $A$ is defining for $c$.
The latter fact, for each $B$, is certified by another proper 3-coloring
$c_B$ that agrees with $c$ on $B$ but differs from $c$ somewhere
outside~$B$.

The completeness is proved by reduction from the decision problem
whether or not $\chi(G)>3$. The latter problem is coNP-complete
even if restricted to connected graphs with $\chi(G)\ge3$. Let $G$ be
a such graph. Our reduction just transforms $G$ into the categorical
product $G\cdot K_3$.
If $\chi(G)>3$, then $G\cdot K_3$ is uniquely 3-colorable
by Lemma \ref{lem:grlo} and, by Lemma \ref{lem:haj},
we have $\fforce{G\cdot K_3}=2$.
If $\chi(G)=3$, then $G\cdot K_3$ has at least two inequivalent
proper 3-colorings, namely, those induced by the two factors.
By Lemma \ref{lem:haj}, we have $\fforce{G\cdot K_3}\ge3$.
\end{proof}

\section{General setting}\label{s:set}

In fact, many other graph characteristics also have natural forcing
variants. Taking those into consideration, it will be convenient to
use the formal concept of an NP optimization problem (see e.g.\ \cite{Cre}).

Let $\{0,1\}^*$ denote the set of binary strings. The length of a string
$w\in\{0,1\}^*$ is denoted by $|w|$.
We will use notation $[n]=\{1,2,\ldots,n\}$.

An {\em NP optimization problem\/} $\pi=(\opt_\pi,\I_\pi,\sol_\pi,\val_\pi)$
(where subscript $\pi$ may be omitted) consists of the following
components.
\begin{itemize}
\item
$\opt\in\{\max,\min\}$ is a {\em type\/} of the problem.
\item
$\I\subseteq\{0,1\}^*$ is the polynomial-time decidable set of
{\em instances\/} of $\pi$.
\item
Given $x\in\I$, we have $\sol(x)\subset\{0,1\}^*$,
the set of {\em feasible solutions\/} of $\pi$ on instance $x$.
We suppose that all $y\in\sol(x)$ have the same length that depends
only on $|x|$ and is bounded by $|x|^{O(1)}$. Given $x$ and $y$,
it is decidable in polynomial time whether $y\in\sol(x)$.
\item
$\val\function{\{0,1\}^*\times\{0,1\}^*}{\bf N}$ is a polynomial-time
computable {\em objective function\/} taking on positive integer values. If
$y\in\sol(x)$, then $\val(x,y)$ is called the {\em value\/} of~$y$.
\end{itemize}
The problem is, given an instance $x$, to compute the optimum value
$$
\pi(x)=\opt_{y\in\sol(x)}\val(x,y).
$$

Such a problem is called {\em polynomially bounded\/} if
$\val(x,y)=|x|^{O(1)}$ for all $x\in\I$ and $y\in\sol(x)$.

Any $y\in\sol(x)$ whose value is optimum is called an {\em optimum
solution\/} of $\pi$ on instance~$x$. Let $\optsol(x)$ denote the set
of all such $y$.
Given an NP optimization problem $\pi$, we define
$$
\UO_\pi=\setdef{x}{|\optsol(x)|=1}.
$$

\begin{example}\label{ex:chi}\rm
The problem of computing the chromating number of a graph is expressible
as a quadruple $\chi=(\min,\I,\sol,\val)$ as follows. A graph $G$ with
vertex set $V(G)=\{v_1,\ldots,v_n\}$ is represented by its adjacency matrix
written down row after row as a binary string $x$ of length $n^2$.
A feasible solution, that is a proper coloring $c\function{V(G)}{[n]}$,
is represented by a binary string $y=c(v_1)\ldots c(v_n)$ of length $n^2$,
where a color $i$ is encoded by string $0^{i-1}10^{n-i}$. The value
$\val(x,y)$ is equal to the actual number of colors occurring in $y$.
\end{example}

\begin{example}\label{ex:omgg}\rm
For the problem of computing the clique number it is natural
to fix the following representation. An instance graph $G$ is encoded as
above. A feasible solution, which is a subset of $V(G)$, is encoded by
its characteristic binary string of length $n$. The problem of computing
the domination number is represented in the same way.
\end{example}

Given a non-empty set $U\subseteq\{0,1\}^l$, we define $\forcef(U)$
to be the minimum cardinality of a set $S\subseteq[l]$ such that there is
exactly one string in $U$ with 1 at every position from $S$.
With each NP optimization problem $\pi$ we associate its {\em forcing
number\/} $F_\pi$, an integer-valued function of instances of $\pi$
defined by
$$
F_\pi(x)=\forcef(\optsol(x)).
$$
Let \decfor\pi{k}$\,=\setdef{x}{F_\pi(x)\le k}$.
It is easy to check that, if $\chi$, $\om$, and $\gh$ are represented
as in Examples \ref{ex:chi} and \ref{ex:omgg}, then $F_\chi$, $F_\om$,
and $F_\gh$ are precisely those graph invariants introduced in
Section \ref{s:intro}.

Note that $\forcef(U)=0$ iff $U$ is a singleton. It follows that
for $\pi\in\{\om,\gh\}$ we have
\begin{equation}\label{eq:uovsf}
x\in\UO_\pi\mbox{\ \ iff\ \ }F_\pi(x)=0.
\end{equation}
This will be the starting point of our analysis of decision problems
\decfor\om{k} and \decfor\gh{k} in the next section.

\section{Hardness of \decfor\om{k} and \decfor\gh{k}}\label{s:omgg}

The results stated here are based on known reducibilities between
several optimization problems related to our work.
Since this material is dispersed through the literature with
proofs sometimes skipped, for the reader's convenience we outline some details.
We first introduce some reducibility concepts for NP optimizations
problems.

Let $\pi$ and $\varpi$ be NP optimization problems of the same type.
Let $f\function{\{0,1\}^*}{\{0,1\}^*}$ and
$g\function{\{0,1\}^*\times\{0,1\}^*}{\{0,1\}^*}$ be polynomial-time
computable functions such that for every $x\in\I_\pi$ we have
$f(x)\in\I_\varpi$ and for every $y\in\sol_\varpi(f(x))$ we have
$g(x,y)\in\sol_\pi(x)$. Such a pair $(f,g)$ is said to be an
{\em $S$-reduction\/} from $\pi$ to $\varpi$ if for
every $x\in\I_\pi$ we have
$$
\opt_\pi(x)=\opt_\varpi(f(x))
$$
and, in addition, for every $y\in\sol_\varpi(f(x))$ we have
$$
\val_\pi(x,g(x,y))=\val_\varpi(f(x),y).
$$

We call an $S$-reduction $(f,g)$ a {\em parsimonious reduction\/} from
$\pi$ to $\varpi$ if, for any $x\in\I_\pi$, $g(x,{\cdot})$ is a one-to-one
map from $\sol_\varpi(f(x))$ onto $\sol_\pi(x)$. If only a weaker
condition is met, namely, that $g(x,{\cdot})$ is a one-to-one
correspondence between the optimum solutions of $\varpi$ on instance $f(x)$
and the optimum solutions of $\pi$ on instance $x$, then $(f,g)$ will
be called a {\em weakly parsimonious reduction\/} from $\pi$ to~$\varpi$.

Given a Boolean formula $\Phi$ in the conjunctive normal form (CNF),
let $\gs(\Phi)$ denote the maximum number of clauses of $\Phi$
satisfiable by the same truth assignment to the variables. By $\gs_3$
we denote the restriction of $\gs$ to 3CNF formulas (those having at most
3 literals per clause). We regard $\gs$ and $\gs_3$ as NP optimization
problems. Both problems belong to the class MAX NP introduced by
Papadimitriou and Yannakakis. Crescenzi, Fiorini, and Silvestri \cite{CFS},
who introduced the notion of an S-reduction, proved that every problem in
MAX NP is $S$-reducible to $\om$, the maximum clique problem. 
We need a somewhat stronger fact about~$\gs_3$.

\begin{lemma}\label{lem:gs3toom}
There is a parsimonious reduction from $\gs_3$ to~$\om$.
\end{lemma}

\begin{proof}
Let $\phi$ be a disjunctive clause and $X$ be the set of variables
occurring in $\phi$. Let $D(\phi)$ denote the set of all conjunctions
that contain every variable from $X$ or its negation and
imply $\phi$. Note that $\phi$ is logically equivalent to
the disjunction of all $\psi$ in $D(\phi)$.

Given a 3CNF formula $\Phi$, we construct a graph $G$ as follows.
Let $V(G)$ be the union of $D(\phi)$ over all clauses $\phi$ of $\Phi$.
We join $\psi_1$ and $\psi_2$ in $V(G)$ by an edge if these conjunctions
are consistent, i.e., no variable occurring in $\psi_1$ occurs in
$\psi_2$ with negation and vice versa.
\end{proof}

\begin{lemma}\label{lem:uoom}
{\bf (Thierauf \cite[Section 3.2.3]{Thi})}
The decision problem $\UO_\om$ is US-hard.
\end{lemma}

\begin{proof}
Denote the restrictions of SAT and USAT to 3CNF formulas by
3SAT and U3SAT respectively. Since there is a parsimonious
$\mone$-reduction from SAT to 3SAT (see e.g.\ \cite{Pap_book}),
U3SAT is US-complete. We now show that $\mbox{U3SAT}\mone\UO_\om$.

Given a 3CNF formula $\Phi$, let $G$ be the graph constructed from $\Phi$
by the reduction of Lemma \ref{lem:gs3toom}. Let $m$ denote the number
of clauses in $\Phi$ and $H=G+2K_{m-1}$, the disjont union of $G$
and two copies of $K_{m-1}$.

If $\Phi\in\mbox{U3SAT}$, then $\om(H)=\om(G)=m$ and $H\in\UO_\om$
because $G\in\UO_\om$.

If $\Phi\in\mbox{SAT}\setminus\mbox{U3SAT}$, then $\om(H)=\om(G)=m$ and
$H\notin\UO_\om$ because $G\notin\UO_\om$.

If $\Phi\notin\mbox{SAT}$, then $\om(H)\le m-1$ and
$H\notin\UO_\om$ having at least two optimal cliques.

Thus, $\Phi\in\mbox{U3SAT}$ iff $H\in\UO_\om$.
\end{proof}

\begin{lemma}\label{lem:omtogg}
$\UO_\om\mone\UO_\gh$.
\end{lemma}

\begin{proof}
Recall that a {\em vertex cover\/} of a graph $G$ is a set $S\subseteq V(G)$
such that every edge of $G$ is incident to a vertex in $S$. The
{\em vertex cover number\/} of $G$ is defined to be the minimum
cardinality of a vertex cover of $G$ and denoted by $\tau(G)$.
It is easy to see and well known that $S\subseteq V(G)$ is a clique in $G$
iff $V(G)\setminus S$ is a vertex cover in the graph complementary to $G$.
It follows that
\begin{equation}\label{eq:uotau}
\UO_\om\mone\UO_\tau.
\end{equation}

We now show that
\begin{equation}\label{eq:uogg}
\UO_\tau\mone\UO_\gh
\end{equation}
by regarding $\tau$ and $\gh$ as NP minimization problems and
designing a weakly parsimonious reduction from $\tau$ to $\gh$.
Recall that, given a set $X$ and a system of its subsets
$\calY=\{Y_1,\ldots,Y_n\}$,
a subsystem $\{Y_{i_1},\ldots,Y_{i_k}\}$ is called a {\em set cover\/} if
$X=\bigcup_{j=1}^kY_{i_j}$. We compose two known reductions between
minimization problems, Reduction A from the minimum vertex cover to
the minimum set cover (\cite[Theorem 10.11]{AHU}) and Reduction B from the
minimum set cover to the minimum domination number (an adaptation of
\cite[Theorem A.1]{Kan}).

{\it Reduction A.}
Given a graph $G$ and its vertex $v$, let $I(v)$ denote the set of
the edges of $G$ incident to $v$. Consider the set $X=E(G)$ and the system
of its subsets $\calY=\setdef{I(v)}{v\in V(G)}$. Then $S\subseteq V(G)$
is an optimal vertex cover for $G$ iff $\setdef{I(v)}{v\in S}$ is an optimal
set cover for $(X,\calY)$.

{\it Reduction B.}
Given a set $X=\{x_1,\ldots,x_m\}$ and a system of sets
$\calY=\{Y_1,\ldots,Y_n\}$ such that $X=\bigcup_{j=1}^nY_j$,
we construct a graph $H$ as follows. $V(H)$ contains each element
$x_i$ in duplicate, namely, $x_i$ itself and its clone $x'_i$.
There is no edge between these $2m$ vertices. Other vertices of $H$
are indices $1,\ldots,n$, with all possible $n\choose2$ edges
between them. If and only if $x_i\in Y_j$, both $x_i$ and $x'_i$
are adjacent to $j$. There are no more vertices and edges.

Observe that any optimal dominating set $D\subset V(H)$ is included
in $[n]$. Indeed, if $D$ contains both $x_i$ and $x'_i$, then it can be
reduced by replacing these two vertices by only one vertex $j$ such that
$x_i\in Y_j$. If $D$ contains exactly one of $x_i$ and $x'_i$, say $x_i$,
then it should contain some $j$ such that $x_i\in Y_j$ to dominate $x'_i$.
But then $D$ can be reduced just by removing $x_i$.

It is also clear that a set $D\subseteq[n]$ is dominating in $H$ iff
$\setdef{Y_j}{j\in D}$ is a set cover for $(X,\calY)$. Thus, there is
a one-to-one correspondence between optimal set covers for $(X,\calY)$
and optimal dominating sets in~$H$.
\end{proof}

Thus, $\UO_\om$ and $\UO_\gh$ are both US-hard.

\begin{lemma}
Let $\pi\in\{\om,\gh\}$. Then $\decfor\pi k\mone\decfor\pi{k+1}$
for any $k\ge0$.
\end{lemma}

\begin{proof}
Given a graph $G$, we have to construct a graph $H$ such that
$F_\pi(G)\le k$ iff $F_\pi(H)\le k+1$. It suffices to ensure that
\begin{equation}\label{eq:hg1}
F_\pi(H)=F_\pi(G)+1.
\end{equation}

Let $\pi=\om$. Let $H$ be the result of adding to $G$ two new vertices
$u$ and $v$ and the edges $\{w,u\}$ and $\{w,v\}$ for all $w\in V(G)$.
Any optimal clique in $H$ consists of an optimal clique in $G$ and of
either $u$ or $v$. Hence any forcing set in $H$ consists of a forcing
set in $G$ and of either $u$ or $v$ (we use the terminology of Section
\ref{s:intro}). This implies~\refeq{eq:hg1}.

If $\pi=\gh$, we obtain $H$ from $G$ by adding a new isolated edge.
\comm{Can $H$ be made connected provided $G$ is such?}
\end{proof}

Putting it all together, we make the following conclusion.

\begin{theorem}
Let $\pi\in\{\om,\gh\}$. Then
$$
\US\mone\UO_\pi=\decfor\pi0\mone\decfor\pi k\mone\decfor\pi{k+1}
$$
for any $k\ge0$.
\end{theorem}

\section{Complexity of \decfor\pi{k}: An upper bound}\label{s:forcekupper}

We first state a simple general property of the class US.

\begin{lemma}\label{lem:usand}
Every US-complete set has an AND$_2$ function.\footnote{%
In fact, a stronger fact is true: every US-complete set has an AND function
of unbounded arity.}
\end{lemma}

\begin{proof}
It suffices to prove the lemma for any particular US-complete set, for
example, \usat. Given two Boolean formulas $\Phi$ and $\Psi$, rename
the variables in $\Psi$ so the formulas become over disjoint sets
of variables and consider the conjunction $\Phi\wedge\Psi$. As easily
seen, the conjunction is in \usat\/ iff both $\Phi$ and $\Psi$ are
in \usat.
\end{proof}

In Section \ref{s:set} with a non-empty set $U\subseteq\{0,1\}^l$ we
associated the number $\forcef(U)$. Additionally, let us put
$\forcef(\emptyset)=\infty$.

\begin{theorem}\label{thm:upper}
Let $\pi$ be a polynomially bounded NP optimization problem.
Then $\decfor\pi k\disj\US$ for each $k\ge0$.
\end{theorem}

\begin{proof}
We will assume that $\pi$ is a minimization problem (the case of
maximization problems is quite similar).
Suppose that $\val(x,y)\le|x|^c$ for a constant $c$.
Given $1\le m\le|x|^c$, we define
$$
\sol^m(x)=\setdef{y\in\sol(x)}{\val(x,y)=m}
$$
and
$$
F_\pi^m(x)=\forcef(\sol^m(x)).
$$
In particular, $F_\pi^m(x)=F_\pi(x)$ if $m=\pi(x)$.

Let $k$ be a fixed integer. Notice that
\begin{equation}\label{eq:iff}
F_\pi(x)\le k\mbox{\ \ iff\ \ }
\bigvee_{m=1}^{|x|^c}\of{
F_\pi^m(x)\le k\wedge\pi(x)\ge m
}
\end{equation}
(actually, only a disjunction member where $m=\pi(x)$ can be true).
The set of pairs $(x,m)$ with $\pi(x)\ge m$ is in coNP and hence in US.
Let us now show that the set of $(x,m)$ with $F_\pi^m(x)\le k$ is
dtt-reducible to US.

Recall that $\sol(x)\subseteq\{0,1\}^{l(x)}$, where $l(x)\le|x|^d$ for a
constant $d$.
Define $T$ to be the set of quadruples $(x,m,l,D)$ such that
$m$ and $l$ are positive integers, $D\subseteq[l]$, and there is a unique
$y\in\sol^m(x)$ of length $l$ with all 1's in positions from $D$.
It is easy to see that $T$ is in US and
$$
F_\pi^m(x)\le k\mbox{\ \ iff\ }
\bigvee_{
\begin{array}{c}
\scriptstyle l,D:\,l\le|x|^d\\[-1mm]
\scriptstyle D\subseteq[l],\, |D|\le k
\end{array}
}
(x,m,l,D)\in T.
$$
Combining this equivalence with \refeq{eq:iff}, we conclude that
$F_\pi(x)\le k$ iff there are numbers $m\le|x|^c$ and $l\le|x|^d$
and a set $D\subseteq[l]$ of size at most $k$ such that
$$
(x,m,l,D)\in T\,\,{}\wedge{}\,\,\pi(x)\ge m.
$$
By Lemma \ref{lem:usand}, this conjunction is expressible as a proposition
about membership of the quadruple $(x,m,l,D)$ in a US-complete set.
Thus, the condition $F_\pi(x)\le k$ is equivalent to
a disjunction of less than $|x|^{c+d(k+1)}$ propositions each verifiable
in~US.
\end{proof}

\begin{corollary}
Let $\pi\in\{\chi,\om,\gh\}$.
Then $\decfor\pi k\disj\US$ for each $k\ge0$.
\end{corollary}

\begin{remark}\rm
Using \refeq{eq:uotau} and Theorem \ref{thm:upper}, we can
easily show that for the vertex cover number $\tau$ we also have
$$
\US\mone\UO_\tau=\decfor\tau0\mone\decfor\tau{k}\mone\decfor\tau{k+1}\disj\US.
$$
\end{remark}

\section{Concluding discussion and open questions}\label{s:open}
\mbox{}

\que
We have considered forcing versions of the three most popular graph invariants:
the chromatic, the clique, and the domination numbers ($F_\chi$, $F_\om$,
and $F_\gh$ respectively). We have shown that the slice decision problems
for each of $F_\chi$, $F_\om$, and $F_\gh$ are as hard as US under
the many-one reducibility and as easy as US under the 
dtt-reducibility. The latter upper bound is actually true
for the forcing variant of any polynomially bounded NP optimization
problem. The lower bound in the case of $F_\om$ and $F_\gh$ is provable
by using standard reductions on the account of a close connection with
the unique optimum problems $\UO_\om$ and $\UO_\gh$. However, in the
case of $F_\chi$ we use somewhat more elaborate reductions involving
graph products. We point out two simple reasons for the distinction
between $F_\chi$ and $F_\om$, $F_\gh$. First,
unlike the case of $\om$ and $\gh$, the unique colorability of a graph
is apparently inexpressible in terms of $F_\chi$
(cf.\ Lemma~\ref{lem:basic}.3). Second,
we currently do not know any reductions between $F_\chi$, $F_\om$, and $F_\gh$
as optimization problems that would allow us to relate their complexities
(cf.\ further discussion).

\que
We have shown that the slice decision problems $\decfor\pi k$ for 
$\pi\in\{\chi,\om,\gh\}$ are close to each other in the complexity hierarchy.
Furthermore, let $\decfor\pi{*}=\setdef{(x,k)}{F_\pi(x)\le k}$.
Hamed Hatami (personal communication) has recently shown that, like 
$\decfor\chi{*}$, the decision problem $\decfor\om{*}$ is $\sitwo$-complete.
Consequently, the problems of computing $F_\chi$ and $F_\om$ are polynomial-time
Turing equivalent. It would be also interesting to
compare the complexities of $F_\chi$, $F_\om$, and $F_\gh$
using weaker reducibility concepts for optimization problems
(as is well known, the similarity of decision versions does not imply
the similarity of the underlying optimization problems; For example,
the decision versions of $\chi$, $\om$, and $\gh$ are all NP-complete
and parsimoniously equivalent but have pairwise different parametrized
complexities and the corresponding optimization problems have
pairwise different approximation complexities).

\que
Is \decfor\pi{k} NP-hard under $\mone$-reductions for any $\pi$
under consideration and constant $k$? It should be noted that the affirmative answer
would settle a long-standing open problem if $\cpclass$ contains NP
in the affirmative.
\comm{Klivans and van Melkebeek "Graph Nonisomorphism Has
Subexponential Size Proofs Unless The Polynomial-Time Hierarchy Collapses"
can derandomize the Valiant-Vazirani reduction under some plausible
hardness assumptions and, under these assumptions, prove that NP is in SPP.
}

\que
Let UCOL be the set of all uniquely colorable graphs (with no restriction on
the chromatic number). Is it true that $\mbox{UCOL}\mone\decfor\chi2$?
\comm{
Note that $\mbox{UCOL}\ne\UO_\chi$ if the latter set is understood
in the formal sense of Section~\ref{s:set}. 
}
It is not hard to show that UCOL is US-hard.
\comm{Indeed, let us show a very simple reduction from U3COL. Given
a graph $G$ with $\chi(G)\ge3$, we choose two adjacent vertices
$u,v\in V(G)$ and let $H$ be the result of adding to $G$ a new vertex $w$
and two new edges $\{u,w\}$ and $\{v,w\}$.
If $G\in\mbox{U3COL}$, then $H\in\mbox{U3COL}\subset\mbox{UCOL}$.
If $G\in\mbox{3COL}\setminus\mbox{U3COL}$, then
$H\in\mbox{3COL}\setminus\mbox{U3COL}$ and hence $H\notin\mbox{UCOL}$.
If $G\notin\mbox{3COL}$, then for any proper coloring of $G$, there are
at least two choices of a color for $w$ and hence $H\notin\mbox{UCOL}$.}

\comm{
\que
Theorem \ref{thm:f2} estimates the relative complexity of the class of
graphs with forcing chromatic number 2 with respect to the wider classes
shown in Figure 2. The picture will be complete if we can estimate
complexity of \ucol\/ under the promise that $\force G=2$.
}

\paragraph{Acknowledgement}

This work was initiated when the authors met at the Department
of Information Systems at Vienna University of Technology.
They are indebted to Georg Gottlob for his hospitality.

\paragraph{In Memoriam}

The journal version of this paper is prepared already without Frank Harary.
He passed away a month before the conference presentation at STACS'05 \cite{stacs}.
Frank's enthusiasm and inspiration were a fundamental of this work.
\hfill {\em W.S.,\ O.V.}

\end{document}